\documentclass[10pt]{article}
\usepackage[margin=1in]{geometry}

\usepackage{url}

\title{Machine Learning over Static and Dynamic Relational Data}

\author{
Ahmet Kara$^1$, 
Milos Nikolic$^2$, 
Dan Olteanu$^1$, 
Haozhe Zhang$^1$ 
\\ \\
$^1$University of Zurich
\enspace\enspace 
$^2$University of Edinburgh
}
\date{}

\begin{document}

\maketitle
\begin{abstract}
This tutorial  overviews principles behind recent works on training  and maintaining
machine  learning models  over relational  data, with an emphasis on
the exploitation of the relational data structure  
to improve the runtime performance of the learning task.  

The tutorial has the following parts:

\begin{enumerate}
\item Database research for data science

\item Three main ideas to achieve performance improvements 
  \begin{enumerate}
  \item[(2.1)] Turn the ML problem into a DB problem
  \item[(2.2)] Exploit structure of the data and problem
  \item[(2.3)] Exploit engineering tools of a DB researcher
  \end{enumerate}

\item Avenues for future research
\end{enumerate}
\end{abstract}

\paragraph{Acknowledgements}
This project has received funding from the European Union's Horizon 2020 research and innovation programme under grant agreement No 682588.

\maketitle

\vspace{5mm}

The starting points of this tutorial are a VLDB 2020
keynote~\cite{DBLP:journals/pvldb/Olteanu20} and a short tutorial at
SUM 2019~\cite{DBLP:conf/sum/SchleichOK0N19}. 
This tutorial goes beyond these prior tutorials, in particular on
learning over dynamic relational data. The interested reader may also want to explore
further work on machine learning over databases overviewed in two
 prior tutorials~\cite{Kumar:SIGMOD:Tutorial:17,Polyzotis:SIGMOD:Tutorial:17}.
 
In the following, we briefly overview each part of the tutorial.

\section{Database research for data science}

Database research is thriving in the data science era. Relational data remains ubiquitous. According to a recent Kaggle survey~\cite{kaggle-survey}, most data scientists use relational data. The widespread use of relational data maintains the relevance of existing relational processing techniques. Furthermore, the new requirements brought by the machine learning workloads have led to new relational processing techniques. In this tutorial, we overview some of these existing and new techniques. They rely to a varying degree on an integration of relational database engines and machine learning libraries.  

This first part of the tutorial overviews two main approaches at the interface of data processing engines and machine learning libraries. The main message is that this interface provides a fruitful and exciting opportunity for database research to shine. A tighter integration of the database and machine learning computation uncovers new research challenges and can lead to significant performance improvements.

A typical approach to learning over relational data involves the construction of the training dataset using a feature extraction query that joins the input relations and constructs new features using aggregates over the data columns. This query can be expressed in SQL and executed using a database system, e.g., PostgreSQL or SparkSQL~\cite{Spark:NSDI:2012}, or Python Pandas~\cite{pandas} for Jupyter notebooks. The desired model is then learned using a machine learning library, e.g., scikit-learn~\cite{scikit2011-small}, R~\cite{R-project}, TensorFlow~\cite{tensorflow-small}, or MLlib~\cite{MLlib:JMLR:2016-small}. This approach ignores the structure of the underlying relational data at the expense of runtime performance. It is dubbed {\em structure-agnostic} in this tutorial. It puts together two black-box specialised systems for data processing and machine learning and may work for virtually any dataset and model. There are several significant downsides of this approach, including: the materialisation of the result of the feature extraction query; the export of this result from the data processing system to the machine learning library; high maintenance cost in case of changes to the underlying data; the limitations of each of the two systems become a limitation of their combination. These downsides hinder the runtime performance of a data science solution using this approach.

The tutorial will also present an alternative approach, dubbed {\em structure-aware} learning. It exploits the data sparsity and the structure of the relational data, in particular the various dependencies in the data and the result of the feature extraction query (multi-valued, functional). In this approach, the learning algorithm is opened up and rewritten such that its data-intensive components are moved closer to the data inside the relational query processor. Such components can be expressed as (group-by) aggregates over the feature extraction query.  Part~\ref{sec:ml2db} of the tutorial is dedicated to them. Their output size is much smaller than that of the feature extraction query and can be computed asymptotically faster than the feature extraction query itself. Since this approach avoids the materialisation of the training dataset and its move between the two systems, it enjoys excellent runtime performance. Opening up the learning black box also allows to consider known mechanisms to maintain the data-intensive components under data updates~\cite{FIVM:SIGMOD:2020}.

The tutorial will contrast the aforementioned two approaches and present experimental evidence for the superior runtime performance of the structure-aware over structure-agnostic~\cite{lmfao}. This runtime performance can in fact be translated into accuracy performance: Within the time budget of training one model with the structure-agnostic approach, the structure-aware approach may train many possible models and eventually choose one with the best accuracy. This part will conclude with a brief account to several instantiations of these two approaches in the literature.

\section{Structure-aware Learning}

Part 1 of the tutorial sets up the scene for Part 2, which overviews  principles behind the structure-aware approach for learning over relational data. We divide this second part of the tutorial into three blocks. The first block looks at how to turn the learning problem into a database problem. 

The second block overviews techniques for computing and maintaining batches of aggregates that arise from mapping the learning problem to a database problem. It considers a variety of types of structure that can be exploited for improving the runtime performance of such techniques, including algebraic, combinatorial, statistical, and geometric structure. Techniques presented in this block aim at lowering the asymptotic computational complexity. 

Finally, the third block overviews engineering tools that proved effective for improving runtime performance, such as parallelisation, specialisation to workload and dataset, and low-level sharing of data access between different components of the code. Such tools can effectively lower the constant factors of the computation time.

\subsection{Turn the ML Problem into a DB Problem}
\label{sec:ml2db}

The tutorial will explain how to express the data-intensive computation of the learning task using various forms of aggregation over the data matrix that are readily supported by database query languages. A particular focus will be on the learning task for ridge linear regression, support vector machines, and tree-structured Bayesi\-an networks (Chow-Liu trees) -- as representatives for wider classes of models and objective functions for the learning task. An in-depth treatment for specific models and objective functions is provided in the literature, e.g.,~\cite{KuNaPa15,SOC:SIGMOD:16,ANNOS:PODS:2018,ANNOS:DEEM:18,lmfao,ANNOS:TODS:2020,faqai,ACMNNOS:TODS:2020}. 

For learning using the least-squares loss function, the gradient vector of this function is built up using sum-product aggregates over the model features and parameters: For each pair of features (database attributes) $x_i$ and $x_j$, there is one such  aggregate \texttt{sum($x_i*x_j$)} that sums over all tuples in the training dataset the product of the values of the two attributes. In case an attribute corresponds to a categorical feature, then it is promoted from the sum to the group-by clause. If both attributes are in the group-by clause, then the sum becomes the count of the number of occurrences of each pair of categories of the two attributes: \texttt{sum($1$) group by $x_i,x_j$}. This approach applies to ridge linear regression~\cite{SOC:SIGMOD:16,ANNOS:PODS:2018} and polynomial regression models in general~\cite{OS:SIGREC:2016}, factorisation machines~\cite{ANNOS:DEEM:18,ANNOS:PODS:2018,ANNOS:TODS:2020}, sum-product networks~\cite{George:thesis:2020}, principal component analysis~\cite{ANNOS:PODS:2018}, quadratically regularised low-rank models~\cite{Gabriel:thesis:2019}, and QR and SVD decompositions~\cite{Bas:thesis:2018}.

For decision trees, the computation of the cost functions for each attribute and condition at a decision tree node can be expressed by a sum-product aggregate with a filter condition. The cost functions used by algorithms such as CART~\cite{cart84} for constructing regression trees rely on aggregates that compute the variance of the response $y$ conditioned on a filter restricting the value of an attribute $x_i$ to be equal, less than, or greater than a given constant (threshold $c_j$): \texttt{VARIANCE}($y$) \texttt{ WHERE } $x_i \texttt{ op } c_j$. For a categorical attribute, the filter condition expresses its membership in a set of possible categories. The thresholds and categories are decided in advance based on the distribution of values for $x_i$. The variance aggregate is expressed using the sum of squares, the square of sum, and the count. For classification trees, the aggregates encode the entropy or the Gini index using group-by counts to compute value frequencies in the data matrix. 

A large class of models, including support vector machines, are trained using sub-gradient descent. They use non-polynomial loss functions, such as (ordinal) hinge, Huber, scalene, and epsilon insensitive, that are defined by multiple cases conditioned on additive inequalities of the form $\sum_i x_i\cdot w_i > c$, where $w_i$ and $c$ are constants and $x_i$ are the features. The efficient computation of aggregates conditioned on additive inequalities  calls for new algorithms beyond the classical ones for theta joins~\cite{faqai,ACMNNOS:TODS:2020}. Similar aggregates are derived for $k$-means clustering~\cite{faqai}.

\subsection{Exploit Structure of Data and Problem}
\label{sec:structure}

This part of the tutorial overviews principles behind new efficient algorithms for batches of group-by aggregates, worst-case optimal equality joins, and additive inequality joins. Such algorithms power the structure-aware learning paradigm by systematically exploiting the structure of the relational data to lower the computational complexity and improve the runtime performance. The tutorial discusses the algebraic, combinatorial, statistical, and geometric structure of relational data, with a focus on the algebraic structure.

{\noindent\bf A. Algebraic structure.} A relation is a sum-product expression, where the sum is the set union and the product is the Cartesian product. The computation expressed using relational algebra can be captured using (semi)rings. There is extensive work in the literature on $k$-relations over provenance semirings~\cite{PROVSEMIRING:PODS:2017}, generalised multiset relations~\cite{DBRING:PODS:2010}, and factors over multiple semirings~\cite{faq}. The tutorial will overview particular properties that make the rings effective for computing and maintaining aggregates, as required by structure-aware learning.

{\noindent\bf A.1. Distributivity of product over sum.} This law allows to factor out data blocks common to several tuples in a relation, represent them once, and compute over them once. It is the main conceptual ingredient of factorised databases~\cite{OlZa15,OS:SIGREC:2016}. By systematically applying the distributivity law, relations can be represented more succinctly yet losslessly as directed acyclic graphs with fewer data value repetitions. The factorisations of relations representing the answers to relational queries can be asymptotically smaller than the standard representation as list of tuples. This applies to a rich class of queries  made up of joins~\cite{FDB:PVLDB:2012,OlZa15}, selections, projections, unions, group-by aggregates~\cite{BKOZ13}, and order-by clauses~\cite{BKOZ13}. More relevant to our tutorial, factorisation can also lower the computational complexity of machine learning over feature extraction queries~\cite{SOC:SIGMOD:16,OS:SIGREC:2016,ANNOS:PODS:2018,faqai}. The tutorial will exemplify factorised computation for joins and aggregates such as those discussed in Part~\ref{sec:ml2db}. The framework of Functional Aggregate Queries~\cite{faq} generalises factorised databases to semirings beyond sum-product and shows that many problems across Computer Science can benefit from factorised computation. LMFAO~\cite{lmfao,LMFAO:PVLDB:2020}, F-IVM~\cite{Nikolic:FIVM:2018, FIVM:SIGMOD:2020}, and IVM$^{\epsilon}$~\cite{IVMe:PODS:2020,IVMe:TODS:2020} employ factorised query computation and maintenance. These algorithms factorise the query into a hierarchy of increasingly simpler views, which are maintained bottom-up under data updates. 

{\noindent\bf A.2. Sum-product abstraction.} By conveniently overloading the sum and product operations in a (semi)ring, we can capture the computation for many different tasks over relational data, including the type of aggregates needed for training models. This is exemplified extensively in the FAQ framework~\cite{faq}, where the same structure of computation but with possibly different definitions of sum and product can be used to compute, among others, database queries, matrix chains, and inference queries (marginals and MAP) in probabilistic graphical models. The tutorial will exemplify this for covariance matrices~\cite{Nikolic:FIVM:2018} and mutual information~\cite{FIVM:SIGMOD:2020} over feature extraction queries, used in the context of training models over relational data. 

{\noindent\bf A.3. Additive inverse.} To accommodate efficient mechanisms for incremental maintenance of learned models over feature extraction queries, we extend tuples to carry along payloads, which are elements from a ring. These payloads can for instance capture the multiplicities of the corresponding tuples or, in case of tuples in the query result, the number of derivations of that tuple from the input tuples via the query. Similarly, the aggregates needed for training a specific model may represent payloads.  Tuple inserts and deletes to the underlying data can be modelled uniformly as inserts with appropriate payloads $a$ and respectively $-a$ that follow the additive inverse law of the ring: $a+ (-a) = 0$, where $0$ is the neutral element for summation in the ring. Whenever a tuple's payload becomes $0$, then it is not anymore part of the data.  In the simplest case, $a$ is an integer representing the multiplicity~\cite{DBRING:PODS:2010,Koch:DBToaster:2014,IVMe:ICDT:2019}. Further examples discussed in the literature consider rings for factorised data representation and covariance matrices~\cite{Nikolic:FIVM:2018,FIVM:SIGMOD:2020}. The efficient maintenance of covariance matrices makes it possible to keep ridge linear regression models fresh under high-throughput data changes~\cite{Nikolic:FIVM:2018}. A recent tutorial overviews advances in incremental view maintenance~\cite{Iman:CIKM:2019}. Our tutorial will go through several examples showing how rings can be used to maintain models under data updates.

{\noindent\bf B. Combinatorial structure.} The combinatorial structure prevalent in relational data is captured by notions such as the width measure of the query and the degree of a data value. For reasons of time limitation, the tutorial will only mention this type of structure. A brief overview is given in a recent keynote~\cite{DBLP:journals/pvldb/Olteanu20}. If a feature extraction query has width $w$, then its data complexity is $\tilde{O}(N^w)$ for a database of size $N$, where $\tilde{O}$ hides logarithmic factors in $N$. Similar complexities have been shown for learning a variety of models over feature extraction queries~\cite{SOC:SIGMOD:16,OS:SIGREC:2016,faqai} using prototypes such as F~\cite{SOC:SIGMOD:16} and LMFAO~\cite{lmfao}. There are several width measures proposed in the literature yet they are beyond the scope of our tutorial. The degree information captures the number of occurrences of a data value in the input database~\cite{skew}. Several existing query processing and maintenance algorithms, e.g., worst-case optimal join algorithms~\cite{NPRR12} and worst-case optimal incremental maintenance of triangle~\cite{IVMe:ICDT:2019,IVMe:TODS:2020}
and hierarchical~\cite{IVMe:PODS:2020} queries, adapt their execution strategy depending on the degree of data values, with different strategies for high-degree and low-degree values.
A special form of bounded degree is given by functional dependencies. They can be used to lower the learning runtime for ridge polynomial regression models and factorisation machines. Instead of learning a given model, we can instead learn a reparameterised model with fewer parameters and then map it back to the original model~\cite{ANNOS:PODS:2018,ANNOS:TODS:2020}.

{\noindent\bf C. Statistical structure.} In case of very large datasets, a feasible approach is to learn approximately over data samples~\cite{murphy2013}. When learning over feature extraction queries, we would like to sample from the input data through the queries. Prior work considered the problem of sampling through selection conditions and joins, e.g., the ripple joins~\cite{Haas:RippleJoin:SIGMOD:1999} and the wander joins~\cite{Li:WanderJoin:TODS:2019}, and for specific classes of machine learning models~\cite{Park:SampleML:SIGMOD:2019}.

{\noindent\bf D. Geometric structure.} This type of structure becomes relevant whenever we use distance measures. Clustering algorithms can exploit such measures, e.g., the optimal transport distance between two probability measures, and the distance-based triangle inequality. The relational $k$-means (Rk-means)~\cite{Rkmeans:AISTATS:2020} is a prime example of structure-aware learning approaches that exploits the geometric structure of the underlying data. It achieves a constant-factor approximation of the $k$-means objective by clustering over a small coreset instead of the full result of the feature extraction query.

\subsection{Engineering Tools of DB Researcher}
\label{sec:engineering}

Towards taming the computational challenge raised by structure-aware learning, a large effort focuses on a toolbox of systems techniques such as: specialisation for workload, data, and hardware; observing the memory hierarchy and blocking operations; distribution and parallelisation. The tutorial will highlight several recent efforts in this space, in particular on compiling the task of learning specific models over feature extraction queries into efficiently executable low-level code. Such techniques can lead to significant runtime performance improvements, as reported for the AC/DC~\cite{IFAQ:CGO:2020}, F-IVM~\cite{Nikolic:FIVM:2018,FIVM:SIGMOD:2020}, LMFAO~\cite{lmfao,LMFAO:PVLDB:2020}, and IFAQ~\cite{IFAQ:CGO:2020,IFAQ:PVLDB:2021} prototypes. The tutorial will make the case for such a compilation approach. These systems are based on prior work on specialisation for queries and database schema~\cite{Neumann:PVLDB:11,legobase_tods,dblablb}. The LMFAO system also systematically  
	shares computation across the batch of aggregates for structure-aware learning and takes advantage of  multi-core CPU architectures for domain and task parallelism.

\section{Future Research}
\label{sec:future}

The tutorial will conclude with reflections on the state of machine learning over relational data and will pinpoint several directions of future research in systems and theory for structure-aware learning. In particular, it will consider questions on the limits of structure-aware learning and how to make it readily useful for practical data science projects. 
It will also give a glimpse of ongoing work of the authors on maintaining machine learning models under updates.


\bibliographystyle{abbrv}
\bibliography{bibtex}

\end{document}